# Impact of Scattering on Phonon Behavior of Alloys

## Abstract

Mina Aziziha[1], Saeed Akbarshahi[1]

[1]Department of Physics & Astronomy, West Virginia University, Morgantown, WV 26506, USA

The lattice dynamics of $Cu_3Au$, $Ni_{70}Pt_{30}$, $Pd_{90}Fe_{10}$, and $Pd_{96}Fe_{04}$ intermetallic is studied using the DFT calculations. We calculated the phonon dispersions and phonon densities of states along two high symmetry paths of the Brillouin zone by Weighted Dynamical Matrix (WDM) approach. We also compared the results with the supercell approach and inelastic neutron scattering. Furthermore, we calculated the impact of mass and force-constant fluctuations on the $Cu_3Au$ and made a comparison with both WDM and supercell approaches results. The averaged first Nearest Neighbor (1NN) force constants between various pairs of atoms in these intermetallic structures are obtained from the WDM approach.

## Introduction

Phonons can contribute to many interesting physical phenomena such as topological insulator, superconductivity, and thermal strength. Still, there is a need to understand the phonon behavior in ordered and disordered crystal systems. Phonon scattering in alloys is essential as alloys are vastly used [1–7], and the study of the scattering and the contributing factors are crucial. The energy dispersion of phonons provides a wide range of information about the dynamical properties of the material, the ordering behavior, phase stability, and elastic properties. It is an essential input in the calculation of thermodynamic properties like the heat capacities, thermal expansion coefficients, transport properties like diffusivity, and quantities like the electron-phonon interactions. The microscopic understanding of the material properties and the distinct phenomena in materials from their lattice dynamics require robust and accurate theoretical tools. For perfect crystals and ordered alloys, the theory of lattice vibrations has been set up on a rigorous basis. However, the same is not valid for substitutional disordered alloys. The presence of disorder results in scattering that not only depends on the impurity concentration but also crucially on both the relative mass and size differences between the constituent atoms. Novel features in the phonon spectra can be expected if the mass and the interatomic force constants of the impurities differ substantially from those of the host material.

In the last decades to enable the simulation of disorders, a significant amount of work has been done to develop effective methods to derive the disorder averaged physical properties. A feasible, effective medium approximation is a coherent potential approximation (CPA) [8] for only the diagonal disorder. To consider the off-diagonal disorder, the itinerant coherent potential approximation (ICPA) is applicable [9]. An essential advantage of the state-of-art ICPA method is that it can provide the exact representation of the disordered force-constants for vibrational systems with the sum rule obeyed. ICPA provides a self-consistent approach in a single-fluctuation approximation.

Aziziha et al.[3,7,10] introduced the weighted dynamical matrix approach, which is based on the rebuilt of the dynamical matrix by averaging the parent compounds mass and force constants. This approach is straightforward to perform and is computationally efficient compared to other effective medium approaches. However, the versatility and pros and cons of this approach are yet to be studied.

In this work, we are calculating the phonon dispersion and frequencies for $Cu_3Au$, $Ni_{70}Pt_{30}$, $Pd_{90}Fe_{10}$, and $Pd_{96}Fe_{04}$ binary alloys using the weighted dynamical matrix (WDM) approach. To better demonstrate the agreement of the result obtained from this approach, we make a comparison to the inelastic neutron scattering experimental data and supercell results. Also, by modifying the WDM, we can consider the mass and force constant fluctuations and are compared with the reported supercell approach [2].

**Computational Details**

We performed the density functional theory calculations [11,12] with a plane-wave basis set, as implemented in the Quantum-Espresso (QE) code [13]. We employed the Perdew–Burke–Ernzerhof generalized gradient approximation (GGA) exchange-correlation functional[14,15] and Optimized Norm-Conserving Vanderbilt Pseudopotential (ONCVPSP)[16]. A variable cell-structure relaxation was performed in QE until the Hellmann-Feynman force and stress are less than 1mRy/Bohr and 0.1 mRy/Bohr [17,18]. The lattice parameters of Ni, Pd, Pt, Cu, Au, and Fe are 3.50, 3.94, 3.96, 4.15, and 3.46 Å, respectively. This is consistent with previous experiments. The relaxed primitive unit cell with cubic structure ($Fm\bar{3}m$) (225) of Ni, Pd, Pt, Cu, Au, Fe is used to construct a 4×4×4 supercell containing 64 atoms. Also, Spin-polarization considered. The relaxation of this Supercell is done with a $6 \times 6 \times 6$ Monkhorst-Pack *k*-point grid [19]. The energy cut-off 110 Ry for wave functions was employed for calculations. To evaluate the force constants using PHONOPY [20] software. In the supercell method for alloys, a finite-size supercell with defects breaks the space group symmetry and leads to a shrinking BZ in reciprocal space. The state-of-the-art band unfolding methods have been developed for electronic problems to recover the phonon spectra within the BZ of the primitive cell [1,21] as well as for phonon problems [2,22]. Here, we use the unfolding program developed by Ikeda et al. to carry out the phonon band unfolding. For the supercell calculation of the Cu-Au, we used from Ikeda et al. data [2].

**Weighted Dynamical Matrix (WDM) Approach**

To find the phonon modes, one needs to construct the dynamical matrix. The dynamical matrix **D**(**q**) at the wavevector **q** is constructed as follows:

$$D_{ii'}^{\alpha\beta}(\mathbf{q}) = \frac{1}{\sqrt{m_i m_{i'}}} \sum_{l'} \Phi_{\alpha\beta}(0i, l'i') \exp[i\mathbf{q} \cdot (\mathbf{r}_{l'i'} - \mathbf{r}_{0i})], \qquad (1)$$

where $m_i$ is the mass of the $i^{\text{th}}$ atom. Phonon frequencies $\omega(\mathbf{q}, \kappa)$ and mode eigenvectors $\chi(\mathbf{q}, \kappa)$ at $\mathbf{q}$, where $\kappa$ is the band index, are obtained by solving the eigenvalue equation:

$$\mathbf{D}(\mathbf{q})\,\chi(\mathbf{q}, \kappa) = [\omega(\mathbf{q}, \kappa)]^2 \chi(\mathbf{q}, \kappa) \,. \tag{2}$$

To calculate the phonon modes for these alloy samples, first, the Hellmann-Feynman forces of the parent structures. The weighted dynamical matrix is constructed as follows:

$$\overline{D}_{ii'}^{\alpha\beta}(\mathbf{q}) = \frac{1}{\sqrt{m_i m_{i'}}} \sum_{l'} \overline{\Phi}_{\alpha\beta}(0i, l'i') \exp[i\mathbf{q} \cdot (\mathbf{r}_{l'i'} - \mathbf{r}_{0i})], \tag{3}$$

**Results and Discussion**

In this section, we compare our proposed weighted dynamical matrix (WDM) approach results with the inelastic neutron scattering experimental results [23–25] for $Cu_3Au$, $Ni_{70}Pt_{30}$, $Pd_{90}Fe_{10}$, and $Pd_{96}Fe_{04}$ alloys. Also, we compare the results of WDM with Supercell computed results for Cu-Au, which are performed and reported by Ikeda et al. [2]. This comparison is made for mass and force-constant fluctuation, mass fluctuation, force-constant fluctuation, and mass, and force-constant averaged cases.

The results obtained for the phonon dispersion curves and phonon density of states along $[0,0,\zeta]$ and $[0,\zeta,\zeta]$ directions are shown in Figure 1(a-d). The WDM computed, and the experimental results of phonon frequencies agree reasonably well both along $[0,0,\zeta]$ and $[0,\zeta,\zeta]$ directions ($\zeta = \frac{\vec{q}}{\vec{q}_{max}}$, $\vec{q}$ is the phonon wavevector). However, WDM calculations show an underestimation in higher frequencies, especially in Longitude modes. It is because the mass and force constants scattering is not considered and are simply averaged. In the higher frequencies, the lighter atoms dominate and averaging on the mass led to an underestimation. Two peaks in the FCC structure indicate the vibrational resonance due to atomic motion. The highest frequency peak in the DOS plot corresponds to the Longitudinal mode where the phonon band is flat, and the lower frequency peak is related to the top of the lower-lying Transverse modes.

Mass and force-constant scatterings play a crucial role in the phonon scattering of the alloys. However, they are not generally considered in the effective medium approaches. Here, we modified our WDM calculations to be able to consider the mass and force constant scatterings for Cu-Au and compared the results with the supercell approach with consideration of mass and force-constant scattering. We can define the mass scattering [26] $C\frac{\Delta m^2}{\bar{m}^2}$ where $C$ and m are concentration and mass, respectively, in contrast, the nature of force constant fluctuations is, in general, much more obscure. The force constant fluctuations are generated from the locally different chemical environments, which can be further modified by the variation of bond lengths (local lattice distortion). The first nearest-neighbor force constants, shown in Table 1, are an order of magnitude larger than those of the further neighbors, so we used them for the comparison purpose. The mass scattering in CuAu is 0.36, and as observed in Figure 2(b), it has a significant impact on the phonon dispersion, and it cannot be neglected. Also, we observe from Table 2 and Figure 2(c) that the second order of force-constant scattering is large. As a result, we consider the scattering factor on the WDM approach, and we observe an excellent agreement with supercell computed results as we can see in Fig 2 (a-d) that the WDM and supercell approach led to same results.

**Conclusions**

We have investigated the lattice dynamics $Cu_3Au$, $Ni_{70}Pt_{30}$, $Pd_{90}Fe_{10}$, and $Pd_{96}Fe_{04}$ intermetallics using first-principles density functional theory. The agreement between WDM, supercell approach, and neutron scattering experiments is good, but it underestimates in higher frequencies. It can be due to lack of exchange-correlation in DFT and averaging on the masses. To overcome this underestimation, we considered the mass and force constant scatterings and compared the modified WDM results with the supercell result of Cu-Au. This is important from the point of view of the feasibility of using *ab initio* ordered alloy force constants to study the disordered CuAu and impact of the mass and force-constant fluctuations on the phonon dispersion. Also, we can see the mass and force-constant fluctuations do not have a significant effect on the phonon dispersion. The results have been analyzed for the nearest-neighbor force constant values. Also, the behavior of force constant versus bond distance follows the expected trend.

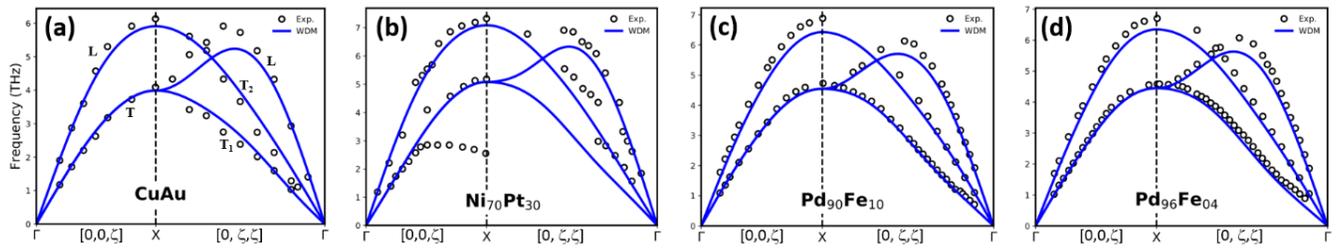

Fig. 1 (a)-(d) Comparison of phonon dispersion of the $Cu_3Au$, $Ni_{70}Pt_{30}$, $Pd_{90}Fe_{10}$ and $Pd_{96}Fe_{04}$ along [0, 0, ζ] and [0, ζ, ζ] directions between WDM approach and experimental results. The blue line in the figures represent the WDM calculations and the circle in the figures represent the experimental results.

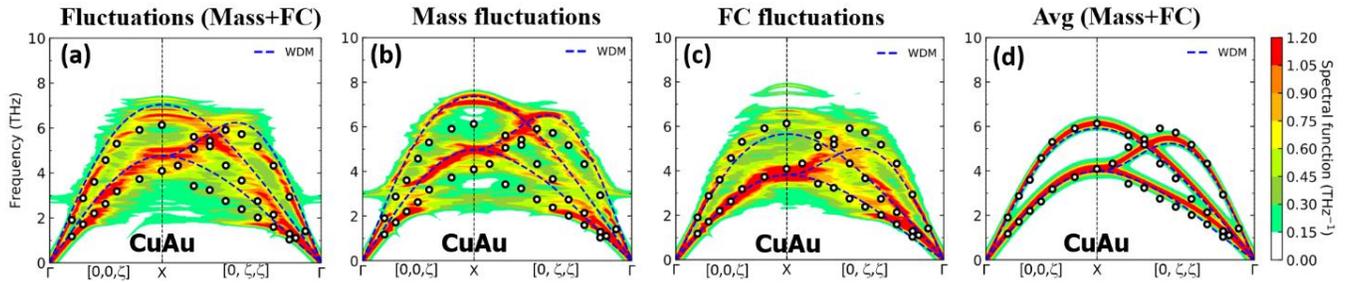

Fig. 2 Comparison of phonon dispersion of CuAu along [0, 0, ζ] and [0, ζ, ζ] directions between Supercell and WDM approach and Impact of mass fluctuations and force constant fluctuations on the phonon dispersion. The blue dashed line in the figures represent the WDM calculations. (a) phonon dispersion including both mass and force constant fluctuations, (b) phonon dispersion including only mass fluctuations, (c) phonon dispersion including only force constant fluctuations, and phonon dispersion including averaged mass and force constants.

Table 1: The averaged first Nearest Neighbor (1NN) force constants, lattice parameter, and bond length in Pt, Ni, Pd, Fe, Cu, Au, CuAu, NiPt, PdFe extracted from supercell calculations. The force constants, lattice parameter, band length, and mass are given in eV/Å$^2$, Å, Å, and AMU, respectively.

|  | Average 1NN $\Phi_{xx,yy,zz}$ (eV/Å$^2$) | Lattice parameter (Å) | Bond length (Å) | Mass (AMU) |
|---|---|---|---|---|
| Pt-Pt | -0.913 | 3.96 | 2.80 | 195.084 |
| Ni-Ni | -0.734 | 3.50 | 2.48 | 58.693 |
| Pd-Pd | -0.699 | 3.94 | 2.79 | 106.420 |
| Fe-Fe | -0.622 | 3.46 | 2.44 | 55.845 |
| Cu-Cu | -0.546 | 3.61 | 2.56 | 63.546 |
| Au-Au | -0.423 | 4.15 | 2.94 | 196.967 |
| CuAu | -0.515 | - | - | 96.901 |
| NiPt | -0.788 | - | - | 99.610 |
| PdFe | -0.691 | - | - | 101.362 |
| PdFe | -0.696 | - | - | 104.396 |

Table 2 The averaged and standard deviations (SD) of the force constants in CuAu, extracted from supercell calculations. The force constants and standard deviations are given in eV/Å$^2$ [2].

|  | Average 1NN $\Phi_{xx}$ | SD | Average 1NN $\Phi_{xy}$ | SD | Average 1NN $\Phi_{zz}$ | SD |
|---|---|---|---|---|---|---|
| Cu-Cu | -0.636 | 0.237 | -0.718 | 0.261 | 0.141 | 0.020 |
| Cu-Au | -1.165 | 0.405 | -1.368 | 0.441 | 0.100 | 0.048 |
| Au-Au | -1.919 | 0.497 | -2.279 | 0.532 | 0.008 | 0.066 |